\documentclass[letterpaper, 10 pt, conference]{ieeeconf}  

\IEEEoverridecommandlockouts                              

\usepackage{graphicx} 
\usepackage{float}

\title{\LARGE \bf
Mining Social Media for Open Innovation in Transportation Systems
}

\author{Daniela Ulloa, Pedro Saleiro, Rosaldo J. F. Rossetti and Elis Regina Silva
\thanks{D. Ulloa is supported by Erasmus Mundus Action 2 Program with an MSc scholarship grant; E. Silva is supported by CNPQ (Conselho Nacional de Desenvolvimento Científico e Tecnológico), Brazil, with doctoral fellowship grant 234933/2014-3.}
\thanks{D. Ulloa, R.J.F. Rossetti, and E. Silva are with the Artificial Intelligence and Computer Science Laboratory (LIACC), Department of Informatics Engineering, Faculty of Engineering, University of Porto, Rua Dr. Roberto Frias, s/n, 4200-465 Porto, Portugal. E-mails: {\tt\small \{up201401125, rossetti, up201409733\}@fe.up.pt.}}%
\thanks{P. Saleiro is with Sapo Labs, Faculty of Engineering, University of Porto, Rua Dr. Roberto Frias, s/n, 4200-465 Porto, Portugal. E-mail: {\tt\small pssc@fe.up.pt}}%
}

\begin{document}

\maketitle
\thispagestyle{empty}
\pagestyle{empty}

\begin{abstract}

This work proposes a novel framework for the development of new products and services in transportation through an open innovation approach based on automatic content analysis of social media data. 
 The framework is able to extract users comments from Online Social Networks (OSN), to process and analyze text through information extraction and sentiment analysis techniques to obtain relevant information about product reception on the market.  
A use case was developed using the mobile application Uber, which is today one of the fastest growing technology companies in the world. We measured how a controversial, highly diffused event influences the volume of tweets about Uber and the perception of its users. While there is no change in the image of Uber, a large increase in the number of tweets mentioning the company is observed, which meant a free and important diffusion of its product.

\end{abstract}

\section{INTRODUCTION}

The connotation of a smart city and a smart transportation system represents high innovation and disruption in management and policy, as well as technology. In the traditional value creation process, companies and consumers had distinct roles of production and consumption. Now, literature about innovation posits that this process lies in an altogether approach to value creation, based on an individual-centered co-creation of value between consumers and companies [1]. Through users’ inputs, it is possible to get improvements on the innovation process because their opinions represent substantial value, bringing benefits for companies that adopt them. Analyzing the field of software development, the involvement of the end user during the development process is not new: there is a need for bringing users’ inputs by collecting their needs through requirements engineering processes, which can be seen as a way of co-creating with users. This process is mainly done through interviews, questionnaires and surveys. Nowadays, this type of data is no longer enough: companies now need access to digital behavioral data as well [2]. With the development of Web 2.0, crowdsourcing emerged as a way of reuniting every possible type of user, allowing a “crowd” of people to collaborate through Web technologies to harness creative collective solutions. While this approach  has been used for long time, it is only recently that businesses and other entities have started increasingly turning to web crowdsourcing as a means of obtaining external expertise [3]. However, both physical and virtual requirements elicitation techniques require manual collection and/or analysis of data, which is costly and time-consuming.

In this regard it is interesting to make a sort of passive crowdsourcing. On the one hand people today comment almost everything about products, releases or news in OSN and on the other technology for word processing on OSN such as sentiment analysis is already available and accessible. Therefore, picking up the available technology and harnessing the huge quantity of data about people’s sentiment, preferences, thinking and opinions available in OSN, we propose a novel framework for the development of new products and services in transportation using an open innovation approach, based on automatic content analysis of social media data. This work follows up previous effort to explore the pontential use of social media in the domain of traffic and transportation engineering [32, 33, 34].

Social media applied to co-creation methods has received an increased deal of attention over the past years, despite this update, the literature is mainly theoretical, needing more empirical research, because the role of social media in companies’ innovation process is not yet clear and its potential on becoming a source of users’ information for firms to improve their NPD process results has not being developed or tested. 

To achieve the objective, we apply an architecture pipeline where we extract users' opinions from the Web, according to previously defined heuristics, for being standardized and having a set of meaningful, general information. Next, using filters, we obtain a knowledge database, where the information gathered before is processed by a natural language processing (NLP), which involves natural language understanding, that is, enabling computers to derive meaning from human or natural language input. Once only data related to the field to be researched is held, we perform quantitative and sentiment analysis processes, with the objective of analyzing the opinion conveyed to each social content, assigning a categorical (positive, negative, neutral) or numerical sentiment score. Finally, the results are shown, through easily understandable ways of visualization.

\section{RELATED WORK}

Three main topics complete the structure of this paper: how the transportation paradigm has changed thanks to the emergence of new needs of transportation technologies in the context of smart cities development; the shift of focus on creating value in companies, seeking more involvement of key actors for an open approach to value creation; and previous research in the area of transportation, smart cities and alike who use text processing of social media.

\subsection{Development of new Transportation Paradigms}

Many studies in the literature have been intensified to introduce new technological solutions in the scope of services in mobility to minimize problems arising from large cities such as traffic congestion, greenhouse gas emissions, fuel consumption, among others [4]. Many works propose models of shared services to make transport services more efficient to both business models and reduction of travel costs for passengers [5]. Shared transportation services that has been most discussed in the literature are Carpooling and Taxi Sharing, these services are aimed at the sharing of a single vehicle for multiple passengers on a journey, and may or may not have local searches and similar destinations [6].

Mobile applications are gaining considerable attention in shared transport services because of their usefulness, not only by the scientific community, but also by many companies. Services such as ``Taxi for Two´´ in United kingdom, where the taxi service is shared by two people who have similar travel routes and share the costs of travel [8]. Another existing service on shared transport is Caronetas-Carpool Smart, whose goal is to share a taxi or a private car to employees or company employees in São Paulo-Brazil [9]. There is also the app ``BlaBlaCar´´ a community of sharing private hitchhikers, present in over 22 countries and with over 30 million members, and allow users to travel by car shares both at national and international level, obtaining economic and social benefits [10].  	

One of the most disruptive applications in the last decade appeared in 2009: Uber, today's one of the most valuable technology companies in the world, the one with fastest growing and the most highly rated of all time. It develops, markets and operates the Uber mobile App, released in June 2010, which allows consumers with smartphones to submit a trip request, which is then routed to Uber drivers who use their own cars. 

\subsection{Open Innovation in the NPD process}

Successful NPD requires two essential types of information: (1) information about customer needs and (2) information about how best to solve these needs [11].

Users have become a valuable source of innovation for companies. Today’s consumers are more connected, informed, empowered and active [12], having different experiences and background, which increase competitive advantages for firms and improve efficiency, turning drive down costs.

Because of their characteristics, companies often take advantage of years of accumulated product knowledge and experience about the precise needs and problems that consumers experience to innovate more successfully [13].
During the years, the degree of involvement of customers have changed, moving from being integrated in the front end innovation (early stages of the process) to a participation in every stage. Given this evolution it is important for companies to change the way they relate with users, shifting their proposition on value creation from a firm-centric view to a customer-centric or co-creation view [12].

Marketing practice and theory increasingly recognize the potential that co-creation has for the firm’s performance [12]. By successfully implementing and managing co-creation, a firm can create two significant sources of competitive advantages [14]: (1) productivity gains through increased efficiency and (2) improved effectiveness.

Co-creation increases productivity and efficiency gains through cost-minimization, since employees’ input can be substituted with consumers’ input in the product/service development [15]. Cost saving arise from various sources: virtually cost-less acquisition of consumer ideas and outsourcing of NPD efforts, which decrease the need for inputs from traditional market research and employees [16], reduced risk of product failure and inventory holding costs [17], faster speed-to-market [18], and post-launch gains through continuous product improvements and exploration of additional usages [19]. These outcomes may directly influence organizational performance, increasing the efficiency of operations, product/service turnover, employee satisfaction and ultimately, revenues and profitability [20].

\subsection{Social Media data processing}

In [26] the authors presented the domain-agnostic framework for intelligent processing of textual streams from social networks. The framework implemented a pipeline of techniques for semantic representation, sentiment analysis, and automatic content classification. The authors used two scenarios related to smart cities to prove the effectiveness of the framework. In the first scenario they monitored the recovery of the state's capital of L'Aquila's city after the earthquake in April 2009, while in the second scenario was performed a semantic analysis of the content posted on the social network Twitter, building an Italian hate map. With the results of the analysis, the authors confirmed that the methodology used for textual content analysis can provide interesting discoveries to improve understanding of the very complex phenomena.

In the study conducted by Saleiro et al. [27] it was presented the POPmine system, an open source system that is able to collect web-based texts from conventional media such as Twitter and blogs and track political views. The authors emphasize that it is the first platform that integrates data collection, information extraction, and opinion mining and public opinion data visualization. The platform processes the text, recognizes the issues and political actors also analyzing relevant language units and generating indicators of dimensions of frequencies and polarities of references to political actors through sources, types of sources and over time.

In [28] TweetCred was developed and evaluated, a real-time web-based system to automatically evaluate the credibility of the Twitter content. The system used a lexicon with 45 feature that included the content of tweets, author of the features and information about URLs. The system was trained from opinions obtained by users on a particular area of study. The system computed rankings credibility of 1.1 million tweets and gave feedback to 936 tweets, of this total 43 of tweets, users agreed with the contents of credibility score calculated by TweetCred. TweetCred was developed in the form of a browser extension, a web application and an API available in http://twitdigest.iiitd.edu.in/TweetCred/.

\section{FRAMEWORK}

We design a platform with the objective of extracting, analyzing and aggregating data in order to produce valuable results by performing four major components: data collection, information extraction, sentiment analysis and analytics. 

The general architecture of the platform is showed in Figure 1. Next, a description of each component is provided. 

   \begin{figure*}[t]
      \centering
      \includegraphics[width=14cm]{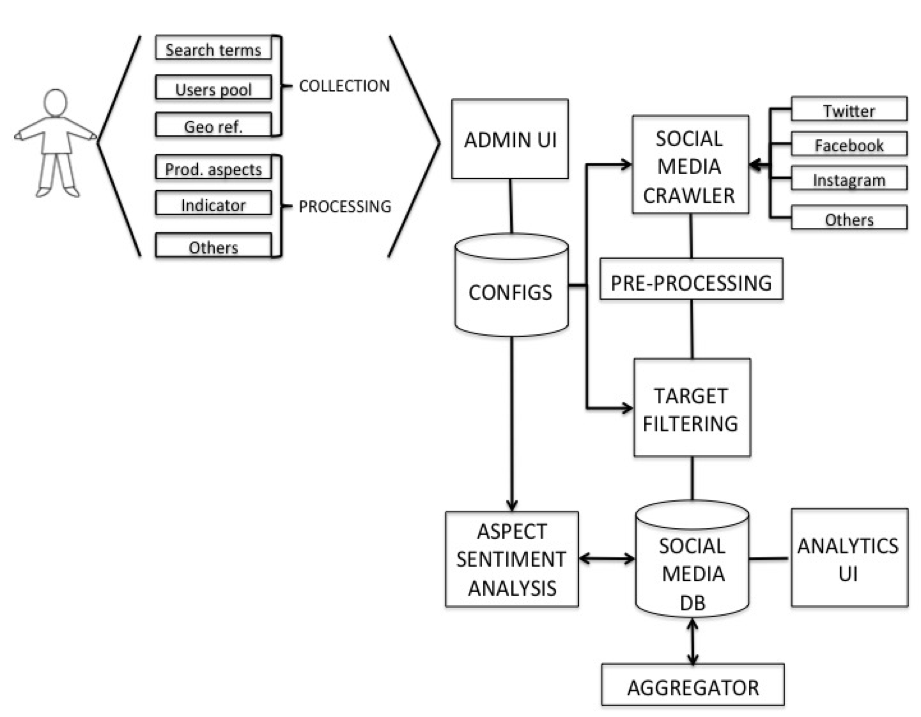}
      \caption{System architecture.}
      \label{figurelabel}
   \end{figure*}
 
\begin{itemize}
\item \textbf{Data collection and processing}
\end{itemize}
For conducting the data collection, the user (company) defines the extraction heuristics, which lead the search and allow connecting to a social network (Facebook, Twitter, Instagram, etc.)  
The data collection components crawl data from specific data sources, which implement specific web interfaces (e.g. RSS feeds, Twitter API). Each data source must have its own data collection module, which in turn connects to the system using REST services. The system stores data collected in a document oriented NoSQL database (MongoDB). This configuration allows modularity and flexibility, allowing the possibility of developing specific data collection components tailored to specific data sources [27]. 

The heuristics defined by the user can be one of three options:
•	Search terms: extracts all data that match a specific term
•	Users pool: extract all data posted by a specific user, given its user name
•	Geo reference: extract all geolocalized data

Next, it is necessary to process the data extracted. For this, the user defines the processing heuristics, which designate specific product’s aspects that are intended to be monitored. Besides, the user also defines the indicators by which the product’s aspects are measured and analyzed.

When having the user’s input, the data collection components crawl data from specific data sources, in our case, a specific OSN, which implement specific web interfaces.

Data are continuously extracted and stored, for the whole time interval an analysis is running. 

\begin{itemize}
\item \textbf{Information extraction}
\end{itemize}
This component addresses two tasks: Named Entity Recognition and Named Entity Disambiguation. This component further processes the content gathered through the data collection process before aggregating, filtering and presenting it in the Analytics UI. This requirement is due to the fact that the extraction process is carried out through a keyword-based matching, so there is a chance that irrelevant content is extracted [26]. When monitoring opinions of a given entity it is first necessary to guarantee that all data are relevant to that entity. 

We envision an application scenario where we need to track specific products entities. Usually this type of entities are well known therefore we opted to use a knowledge base to provide metadata about the target entities, namely the most common surface forms of their names. Once we had the list of surface forms to search for we applied a sequential classification approach using a prefix tree to detect mentions. This method can result in noisy mentions when applied to Social Media. For instance, an opinion containing the Word “Apple” can be related with more than one entity, such as the technology company or a fruit. Furthermore, posts in OSN are usually short which results in a reduced context for entity disambiguation. When monitoring the opinion of a given entity on OSN, it is first necessary to guarantee that all posts are relevant to that entity. 

It is very important to conduct this part of the framework in order to obtain a richer and fine-grained semantic representation, which is very valuable and useful for final users, in virtue of its transparency and readability. Once the filtering process is done, the final data is stored into the Social Media Database [29].

Consequently, other processing tasks, such as sentiment analysis will benefit from filtering out noise in the data stream. 

\begin{itemize}
\item \textbf{Sentiment analysis}
\end{itemize}
It is the computational study of people’s opinions, appraisals, attitudes, and emotions toward entities such as products, services, organizations, individuals, events, and their different aspects [30]. 

Its basic task is to extract and summarize people’s opinions expressed on entities and aspects of entities. It consists of three core sub-tasks:

\begin{itemize}
\item Identifying and extracting entities in evaluative texts

\item Identifying and extracting aspects of the entities

\item Determining sentiment polarities on entities and aspects of entities
\end{itemize}

To this aim, we implement a lexicon-based algorithm for sentiment analysis. Lexicon-based approaches [31] infer the sentiment conveyed by a piece of text by relying on (external) lexical resources, which map each term to a categorical (positive, negative, neutral) or numerical sentiment score. 

We built a sentiment analysis module to detect and classify opinionated posts, mentioning at least one entity, as expressing a positive, negative or neutral opinion. This module was built with a supervised text classification approach based on the bag-of-words assumption, that is, ignoring word order and representing messages as high-dimensional feature vectors with the size of the vocabulary. Then, a manually labeled dataset was used to train a linear classifier that estimates the probability of a sentiment label, given a message. 

One of the main challenges of developing text classification systems for social media, is dealing with the small size of the messages and the large vocabularies needed to account for the significant lexical variation caused by the informal nature of the text. This causes feature vectors to become very sparse, hampering the generalization of the classifiers. Therefore, we first normalized the messages and then enriched the bag-of-words model with additional features based on dense word representations. These word representations, were derived from an unlabeled corpus of 10 Million tweets using two well- known unsupervised feature learning methods: (i) Brown (Brown et al., 1992) hierarchical clustering algorithm, that groups together words that tend to appear in the same contexts. Words are thus represented as the cluster to which they belong; (ii) Mikolov (Mikolov et al., 2013) neural language models that induce dense word vectors (commonly referred to as word embeddings) by training a neural network to maximize the probability that words within a given window size are predicted correctly. 

This component enriches the comprehension of the content by analyzing the opinion conveyed by each social content. With this, we aim to determine the attitude of a user with respect to the aspects indicators previously defined.

\begin{itemize}
\item \textbf{Analytics}
\end{itemize}
A set of analysis can be executed with our architecture. This component let the user visualize and interact with the results of the analysis.

\begin{itemize}
\item Mentions to mobility mobile Apps.

\item Sentiment conveyed through opinions regarding mobility mobile Apps aspects.

\item Evaluation of the performance of mentioned mobility mobile Apps aspects.

\item Following the mapping from section 3.1 determine which innovation characteristics the mentioned mobility mobile Apps match.
\end{itemize}

Through the results the framework provides, companies can improve their NPD process to have clarity about the perception that users have of their products, and what are the ideas for improvement that they propose on the Web, serving as background for companies’ decision making.

\section{USE CASE}

As exemplification of the methodology presented, a use case is developed, taking the mobile App Uber. The aim is that the proposed framework is comprehensible, using a popular and known mobile App that has come in a disruptive way, to change the paradigm of transportation systems.  

Uber Technologies Inc. was founded in 2009 in San Francisco, USA. It develops, markets and operates the Uber mobile App, released in June 2010, which allows consumers with smartphones to submit a trip request, which is then routed to Uber drivers who use their own cars. As of June 3rd, 2016 the service is available in over 66 countries and 469 cities worldwide.

As in other countries, the arrival of Uber to Portugal has not been without controversy. While users advocate its use indicating that the advantages and ease far outweigh traditional taxis, the latter have not received well the arrival of the mobile App, considering it an unfair competition. On April 29, 2016, about four thousand taxis took to the streets of Lisbon in a slow march to protest Uber. The manifestation began at 9:00 am at Park of Nations, and ended after 13:00 at the Parliament. At the same time there were manifestations also in Porto, where about 1500 taxis participated, and in Faro, where adherence was around 500 taxi drivers.
   \begin{figure}[t]
      \centering
      \includegraphics[width=8.3cm]	
      {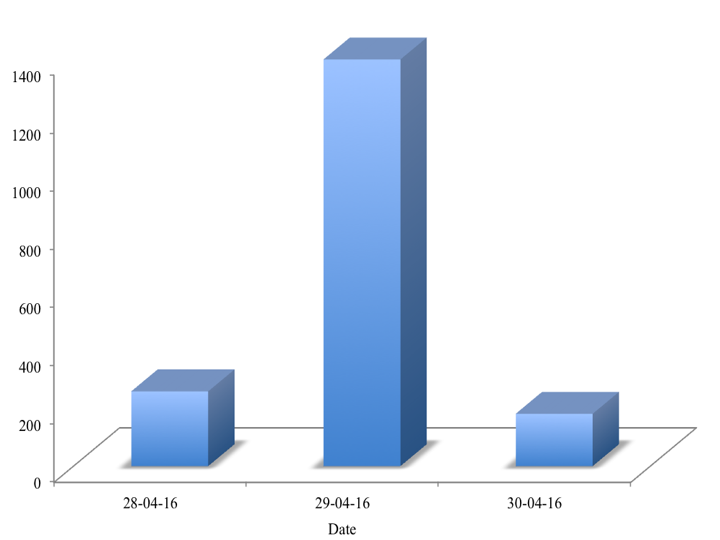}
      \caption{Number of tweets for entity "Uber"}
      \label{figurelabel}
   \end{figure}
The first filter applied was the search entity "Uber". To develop the use case we decided to use tweets generated during manifestation’s previous and next day, so a date filter is established, taking the tweets of the day before and the day after the protest: April 28th (247 tweets) and 30th (180 tweets), 2016. The purpose of this is to analyze, through sentiment analysis, how the protest influenced Uber perception in the population, by comparing the sentiment attached on the tweets before the protest and after the protest. When there are situations of social revolt, also publicized by the media, people tend to react more in OSN, expressing opinions and thoughts; mainly on Twitter, which has become a tool that actually measures the pulse of humanity: hundreds of global social movements are instantly organized and shared through this social network, influencing the perception about the discussed topic.

The whole atmosphere surrounding the demonstration of taxi drivers against Uber in Portugal had a major impact on OSN, reflected on Twitter in a significant increase in entries and comments.

While the data shows that, during that day, the brand Uber became clearly more visible in OSN (Figure 2), the sentiment of users about the brand practically did not changed. While the positive and negative feelings decreased by 3 and 1 percentage point, respectively, the neutral sentiment increased by 3 percentage points (Figure 3 and Figure 4). 

   \begin{figure}[h]
      \centering
      \includegraphics[width=8.3cm]{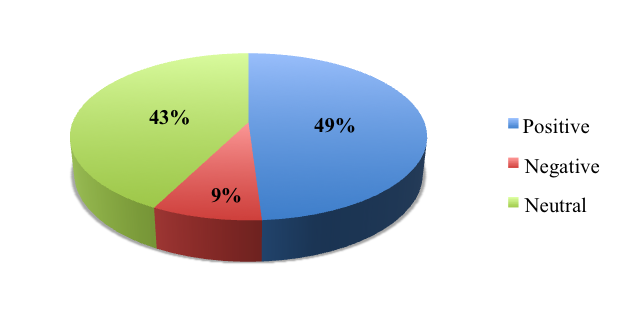}
      \caption{Percentages of tweets for every polarity during April 28th, 2016}
      \label{figurelabel}
   \end{figure}
   
   \begin{figure}[h]
      \centering
      \includegraphics[width=8.3cm]{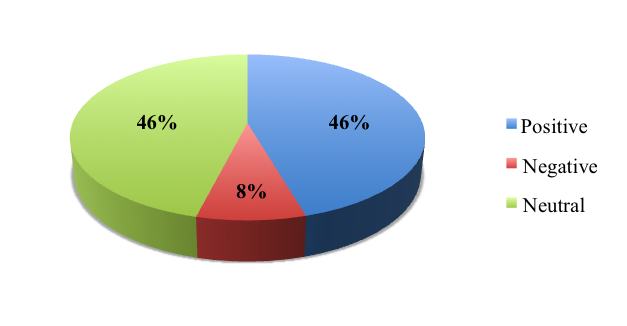}
      \caption{Percentages of tweets for every polarity during April 30th, 2016}
      \label{figurelabel}
   \end{figure}
A massive and opinion-conflicting event, which was highly diffused through OSN and media influenced greatly the visibility of the main matter and installed the brand in the minds of users; but it did not influence the perception that users had about it, because to have enough arguments to change that view requires a deeper process of experimentation and a longer period of time. The main topic of discussion was not focused on functional characteristics of Uber, but rather in its dispute with traditional taxi drivers, so the discussion that took place on Twitter had more to do with this than with an assessment about quality service of Uber. In fact, the aim of the taxi drivers was damaging the image of Uber, but they did not achieve it. This event meant a lot of free publicity for Uber.

\section{CONCLUSIONS AND FUTURE WORK}

The objective of this paper was to assess the impact of OSN as a user-centered information provider for companies to improve their NPD processes, proposing a framework for companies, specifically in the transportation area, to harvest information provided by users. The purpose of creating a system with this information is to support the companies' decision-making process when taking into account users' opinions and sentiment about products, for improving their qualities and better meet users' needs.

As exemplification of the methodology proposed, we developed a use case taking as example the mobile App Uber, chosen for being a disruptive technology, which has quickly become popular in over 66 countries around the world, transforming the traditional transportation concept. To collect the data we applied two filters: the search entity ''Uber´´ and the geolocation “Portugal”. 

In order to evaluate the polarity shift of the population in terms of the perception of Uber because of a protest by traditional taxi drivers, we considered the tweets containing the word “Uber” generated during the days before and after the protest. While there was a significant increase in entries and comments on Twitter about Uber, the sentiment of users about the brand practically did not change. The aim of the taxi drivers was damaging the image of Uber, but this did not happen, in fact, this event meant a lot of free publicity for Uber. A massive, highly diffused and opinion-conflicting event influenced greatly the visibility of the matter under discussion; but it did not influence the perception that users had about it, because to have enough arguments to change that view requires a deeper process of experimentation and a longer period of time. 

The results of the use case are useful for Uber and taxis, maybe especially up to taxis. In addition to the framework be used to evaluate the products, it can also be used to monitor what is said about competitors. And in this case, people/users in Portugal do not discuss the question of the legality of the Uber but above all the advantages of its service. Perhaps taxi drivers association should use these results to try to improve the services provided by taxis, taking into account what people value about Uber.

We believe this research can be deepen by developing the platform for actually having a system to constantly monitor what user express about a product on the Web. It is also important to validate the framework in a closer to reality environment, using actual cases of product development. To this end, next steps in this research include coupling information from social networks to the MAS-Ter Lab framework \cite{c35, c36}, so as to foster and enhance the analysis of new transportation paradigms, in which social simulation and artificial societies are underlying concepts \cite{c37}.

\addtolength{\textheight}{-12cm}   





\end{document}